\documentclass[showpacs,preprintnumbers, nofootinbib]{revtex4}
\usepackage{amssymb}

\usepackage{fontenc}
\usepackage{graphicx}
\usepackage[dvips]{hyperref}
\usepackage{amsmath}

\begin{document}

\title{Quasinormal Modes, Stability Analysis and Absorption Cross Section for $4$-dimensional Topological Lifshitz Black Hole}
\author{P. A. Gonz\'{a}lez}
\email{pgonzalezm@ucentral.cl}
\affiliation{Escuela de Ingenier\'{\i}a Civil en Obras Civiles. Facultad de Ciencias F%
\'{\i}sicas y Matem\'{a}ticas, Universidad Central de Chile, Avenida Santa
Isabel 1186, Santiago, Chile}
\affiliation{Universidad Diego Portales, Casilla 298-V, Santiago, Chile.}
\author{Felipe Moncada}
\email{fmoncada@ufromail.cl}
\affiliation{Departamento de Ciencias F\'{\i}sicas, Facultad de
Ingenier\'{i}a, Ciencias
y Administraci\'{o}n, Universidad de La Frontera, Avenida Francisco Salazar
01145, Casilla 54-D, Temuco, Chile.}
\author{Yerko V\'{a}squez}
\email{yvasquez@ufro.cl}
\affiliation{Departamento de Ciencias F\'{\i}sicas, Facultad de
Ingenier\'{i}a, Ciencias
y Administraci\'{o}n, Universidad de La Frontera, Avenida Francisco Salazar
01145, Casilla 54-D, Temuco, Chile.}

\date{\today}





\begin{abstract}
We study scalar perturbations in the background of a Topological Lifshitz black hole in four dimensions. We compute analytically the quasinormal modes and from these modes we show that Topological Lifshitz black hole is stable. On the other hand, we compute the reflection and transmission coefficients and the absorption cross section and we show that there is a range of modes with high angular momentum which contributes to the absorption cross section in the low frequency limit. Furthermore, in this limit, we show that the absorption cross section decreases if the scalar field mass increases, for a real scalar field mass.
\end{abstract}
%
%
\maketitle
\section{Introduction}
\label{intro}
The Lifshitz black holes are interesting because they are solutions of theories that exhibit the anisotropic scale invariance $t\rightarrow \lambda ^z t$, $x\rightarrow \lambda x$, where $z$ is the relative scale dimension of time and space\footnote{ If $z=1$, the spacetime is the usual AdS metric in Poincare coordinates} and by its representation as the gravitational dual of strange metals~\cite{Hartnoll:2009sz}. In the present work we consider the four dimensional Topological Lifshitz black holes~\cite{Mann:2009yx}, with dynamical exponent $z=2$, which is described by the metric 
\begin{equation}\label{metric1}
ds^2=-\frac{r^2}{l^2}\left(\frac{r^2}{l^2}-\frac{1}{2}\right)dt^2+\frac{dr^2}{\left(\frac{r^2}{l^2}-\frac{1}{2}\right)}+r^2d\Omega^2~,
\end{equation}
where $d\Omega^2=d\theta^2+sinh^2\theta d\phi^2$, and we present exact solutions for the quasinormal modes (QNMs), the reflection and transmission coefficients and the absorption cross section. The QNMs, as the absorption cross section have a long history, see \cite{Regge:1957td},  \cite{Zerilli:1971wd}, \cite{Zerilli:1970se}, \cite{Kokkotas:1999bd},  \cite{Nollert:1999ji},  \cite{Berti:2009kk},  \cite{Konoplya:2011qq} and \cite{Starobinsky},  \cite{Gibbons:1975kk},  \cite{Page:1976df},  \cite{Unruh:1976fm},  \cite{Dhar:1996vu},  \cite{Kol:1996hf},  \cite{Gibbons:1993hg}, respectively. 
The QNMs have been mainly related to the quantized black hole area and mass \cite{Hod:1998vk},  \cite{Kunstatter:2002pj} and \cite{Maggiore:2007nq}. Furthermore, the QNMs have been also related to thermal conformal field theories \cite{Horowitz:1999jd} in the context of AdS/CFT
correspondence \cite{Maldacena:1997re}. In \cite{Birmingham:2001pj} a quantitative test of such a correspondence was
made for the BTZ black hole. Another consequence of AdS/CFT correspondence
has been the application beyond high energy physics to another areas of
physics, for example quantum chromodynamics and condensed matter. In this
sense, Lifshitz spacetimes have received great attention from the condensed
matter point of view i.e., the searching of gravity duals of Lifshitz fixed
points. In \cite{Kachru:2008yh} has been conjectured gravity duals of non
relativistic Lifshitz-like fixed points. This correspondence have been also
studied in \cite{Hartnoll:2009sz}. In this paper we study the quasinormal modes of scalar
perturbations, therefore through this correspondence the quasinormal modes
of Lifshitz black holes have an interpretation as the poles of the
holographic Green's function related to the bulk scalar field.

On the other hand, the absorption cross section allows us to know the absorption probability and to study the scattering problem. Recently, the QNMs have been reported, as well as the stability analysis and the area spectrum of three dimensional Lifshitz black holes~\cite{CuadrosMelgar:2011up} and the absorption cross section~\cite{Moon:2012dy} and  \cite{Lepe:2012zf}, the QNMs and stability analysis of four dimensional Lifshitz black hole with plane transverse section~\cite{Gonzalez:2012de} and  \cite{Giacomini:2012hg} and the QNMs and thermodynamic quantities for three and four dimensional Lifshitz black holes~\cite{Myung:2012cb}. 

In this work, we analize numerically the behavior of the reflection and the transmission coefficients and the absorption cross section of four dimensional Topological Lifshitz black holes and we find that there is a range of modes with high angular momentum which contributes to the absorption cross section and that it decreases if the scalar field mass increases, for real scalar field masses in the low frequency limit. 

This paper is organized as follows. In Sec. \ref{QNM}, we consider Dirichlet and Neumann boundary conditions and we calculate the
exact QNMs of the scalar perturbations. Then, from these quasinormal modes, we study the stability of this black hole. In Sec. \ref{coefficients} we calculate the reflection and the transmission coefficients and the absorption cross section. Finally, our conclusions are
in Sec. \ref{remarks}.
\section{Quasinormal Modes}
\label{QNM}
The QNMs of scalar perturbations in the background of a Topological Lifshitz black hole in four dimensions are given by Klein-Gordon equation of the scalar field solution with 
boundary conditions. This means that there is only ingoing waves on the horizon and the scalar field, Dirichlet boundary condition, or the flux, Neumann boundary condition,  vanishes at infinity. The Klein-Gordon equation is
\begin{equation}\label{KG}
\frac{1}{\sqrt{-g}}\partial_{\mu}\left(\sqrt{-g} g^{\mu\nu}\partial_{\nu}\right)\psi=m^2\psi~,
\end{equation}
where $m$ is the mass of the scalar field $\psi$, which is minimally coupled to curvature.
In order to find an exact solution to the scalar field we will perform the following change of variables $y=\frac{2r^2}{l^2}$, which will allow us to write the metric (\ref{metric1}), as
\begin{equation}\label{metric3}
ds^2=-\frac{y}{4}\left(y-1\right)dt^2+\frac{l^2dy^2}{4y\left(y-1\right)}+\frac{l^2y}{2}d\Omega^2~.
\end{equation}
Now, if we consider the ansatz $\psi=R(y)Y(\sum)e^{-i\omega t}$, where $Y$
is a normalizable harmonic function on $\sum_{d-2}$ 
which satisfies $\nabla^2Y=-QY$, being  
\begin{equation}
Q=\left(\frac{d-3}{2}\right)^2+\xi^2,
\end{equation}
the eigenvalues for the hyperbolic manifold. 
The Klein-Gordon equation~(\ref{KG}), can be written as 
\begin{eqnarray}\label{equationy}
 y\left(1-y\right)\partial_{y}^2R(y)+\left(2-3y\right)\partial_{y}R(y)+\left[\frac{l^2\omega^2}{1-y}+\left(l^2\omega^2+\frac{Q}{2}\right)\frac{1}{y}+\frac{m^2l^2}{4}\right]R(y)=0~.
\end{eqnarray}
If we define $z=\frac{y-1}{y}$ the equation~(\ref{equationy}), becomes 
\begin{eqnarray}\label{radial1}
z\left(1-z\right)\partial_{z}^2R(z)+\partial_{z}R(z)+\left[\frac{\omega^2 l^2}{z}-\left(l^2\omega^2+\frac{Q}{2}\right)-\frac{m^2l^2}{4(1-z)}\right]R(z)=0~,
\end{eqnarray}
and the decomposition $R(z)=z^\alpha(1-z)^\beta K(z)$, with
\begin{equation}
\alpha_{\pm}=\pm i \omega l~,
\end{equation}
\begin{equation}\label{omega2}\
\beta_{\pm}= \frac{1}{2}\left(2\pm \sqrt{4+ m^2l^2}\right)~,
\end{equation}
let us to write (\ref{radial1}), as a hypergeometric equation for K
\begin{eqnarray}\label{hypergeometric1}\
 z(1-z)K''(z)+\left[c-(1+a+b)z\right]K'(z)-ab K(z)=0~,
\end{eqnarray}
where the coefficients are given by
\begin{equation}\label{a}\
a=\frac{1}{2}\left(-1+2\alpha+2\beta\pm\sqrt{1-2Q-4\omega^2l^2}\right)~,
\end{equation}
\begin{equation}
b=\frac{1}{2}\left(-1+2\alpha+2\beta\mp\sqrt{1-2Q-4\omega^2l^2}\right)~,
\end{equation}
\begin{equation}
c=1+2\alpha~.
\end{equation}
The general solution of the hypergeometric equation~(\ref{hypergeometric1}) is
\begin{eqnarray}
K=C_{1}F_{1}(a,b,c;z) +C_2z^{1-c}F_{1}(a-c+1,b-c+1,2-c;z)~,
\end{eqnarray}
and it has three regular singular points at $z=0$, $z=1$. Where 
$z=\infty$. $F_{1}(a,b,c;z)$ is a hypergeometric function
and $C_{1}$ and $C_{2}$ are constants. Then, the solution for the
radial function $R(z)$ is
\begin{eqnarray}\label{RV}\
R(z)=C_{1}z^\alpha(1-z)^\beta
F_{1}(a,b,c;z)+C_2z^{-\alpha}(1-z)^\beta
F_{1}(a-c+1,b-c+1,2-c;z)~.
\end{eqnarray}
So, in the vicinity of the horizon, $z=0$ and using
the property $F(a,b,c,0)=1$, the function $R(z)$ behaves as
\begin{equation}\label{Rhorizon}
R(z)=C_1 e^{\alpha \ln z}+C_2 e^{-\alpha \ln z},
\end{equation}
and the scalar field $\psi$, for $\alpha=\alpha_-$, can be written in the following way
\begin{equation}
\psi\sim C_1 e^{-i\omega(t+ l\ln z)}+C_2
e^{-i\omega(t-l\ln z)}~,
\end{equation}
in which, the first term represents an ingoing wave and the second
one an outgoing wave in the black hole. So, by imposing that
 only ingoing waves exist at the horizon, this fixes  $C_2=0$. Then, the radial
solution becomes
 \begin{eqnarray}\label{horizonsolution}
R(z)=C_1 e^{\alpha \ln z}(1-z)^\beta F_{1}(a,b,c;z)=C_1e^{-i\omega l\ln z}(1-z)^\beta F_{1}(a,b,c;z)~.
\end{eqnarray}
To implement boundary conditions at infinity ($z=1$), we
shall apply the Kummer's formula
for the hypergeometric function \cite{M. Abramowitz},
\begin{eqnarray}\label{relation}
F_{1}(a,b,c;z)=\frac{\Gamma(c)\Gamma(c-a-b)}{\Gamma(c-a)\Gamma(c-b)}F_1
 +(1-z)^{c-a-b}\frac{\Gamma(c)\Gamma(a+b-c)}{\Gamma(a)\Gamma(b)}F_2~.
\end{eqnarray}
Where,
\begin{equation}
F_1=F_1(a,b,a+b-c,1-z)~, 
\end{equation}
\begin{equation}
F_2=F_1(c-a,c-b,c-a-b+1,1-z)~.
\end{equation}
With this expression the radial function~(\ref{horizonsolution}) reads
\begin{eqnarray}\label{R}\
R(z) = C_1 e^{-i\omega l \ln z}(1-z)^\beta\frac{\Gamma(c)\Gamma(c-a-b)}{\Gamma(c-a)\Gamma(c-b)} F_1+C_1 e^{-i\omega l \ln
z}(1-z)^{c-a-b+\beta}\frac{\Gamma(c)\Gamma(a+b-c)}{\Gamma(a)\Gamma(b)}F_2~,
\end{eqnarray}
and at infinity can be written as 
\begin{eqnarray}\label{R2}\
R_{asymp.}(z) = C_1 (1-z)^\beta\frac{\Gamma(c)\Gamma(c-a-b)}{\Gamma(c-a)\Gamma(c-b)}+C_1 (1-z)^{2-\beta}\frac{\Gamma(c)\Gamma(a+b-c)}{\Gamma(a)\Gamma(b)}~.
\end{eqnarray}

For,  $\beta_+>2$ the field at infinity vanishes
if $(a)|_{\beta_+}=-n$ or $(b)|_{\beta_+}=-n$ for $n=0,1,2,...$. For,  $\beta_-<0$ the field at infinity vanishes 
if $(c-a)|_{\beta_-}=-n$ or $(c-b)|_{\beta_-}=-n$. Therefore, the quasinormal modes are given by
\begin{eqnarray}\label{w1}
\omega_1=-i \frac{\sqrt{4+m^2l^2}+2n}{4l}-i \frac{2Q+\sqrt{4+m^2l^2}+2n}{4l\left(\sqrt{4+m^2l^2}+2n+1\right)}~.
\end{eqnarray}
Note that four dimensional Topological Lifshitz black hole is stable due to the imaginary part of the QNMs~(\ref{w1}), is negative. 

Now, we consider that the flux
vanishes at infinity (or vanishing Neumann boundary condition at infinity) in order to find the QNMs for imaginary masses. Because in asymptotically Lifshitz spacetime, a negative square mass for a
scalar field is consistent, in agreement with the analogue to the
Breitenlohner-Freedman condition that any effective mass must satisfy in
order to have a stable propagation, \cite{Breitenlohner:1982bm},
\cite{Breitenlohner:1982jf}. Dirichlet boundary condition must leads to the same
quasinormal modes for positive square mass from a holographic point of view,
but does not lead to any quasinormal modes for negative square mass.
However, quasinormal modes can also be obtained by considering Neumann
boundary conditions and in the context of conformal field theory, due to the
 AdS/CFT correspondence \cite{Maldacena:1997re}, the quasinormal modes match
with the dual operators, \cite{Birmingham:2001pj}.

\begin{figure}
\includegraphics[width=0.45\textwidth]{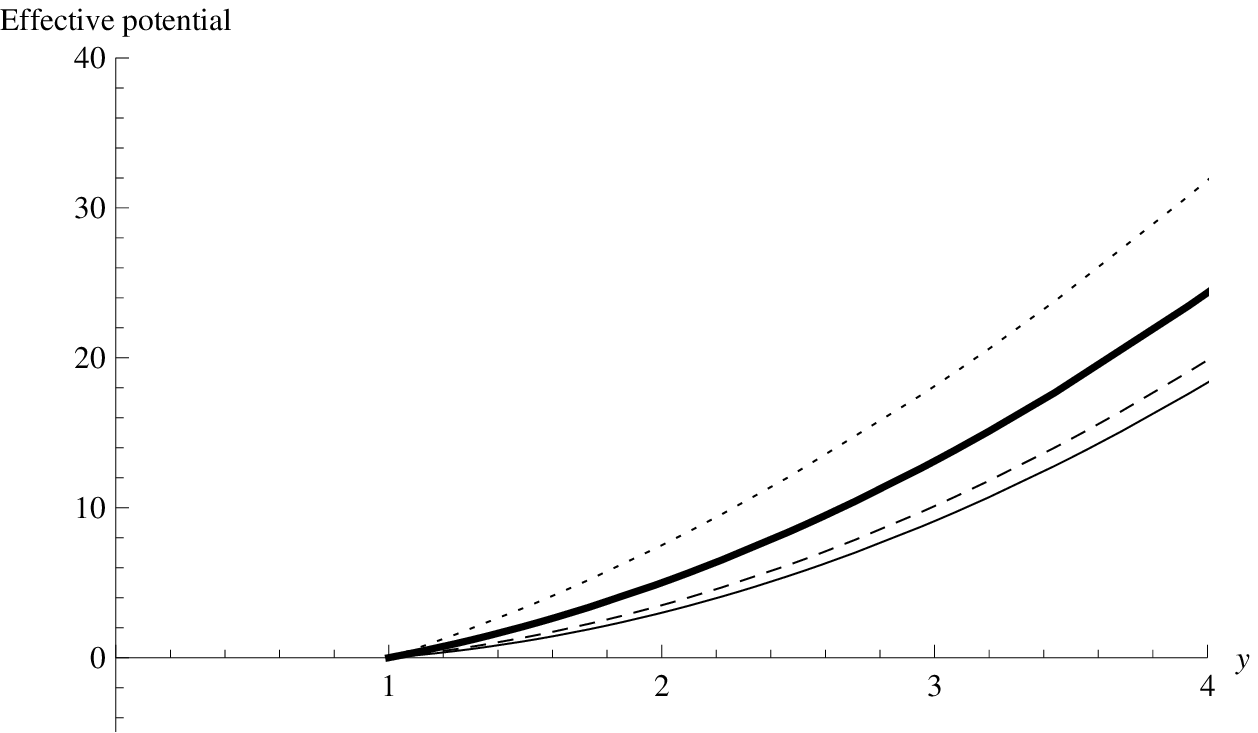}
\includegraphics[width=0.45\textwidth]{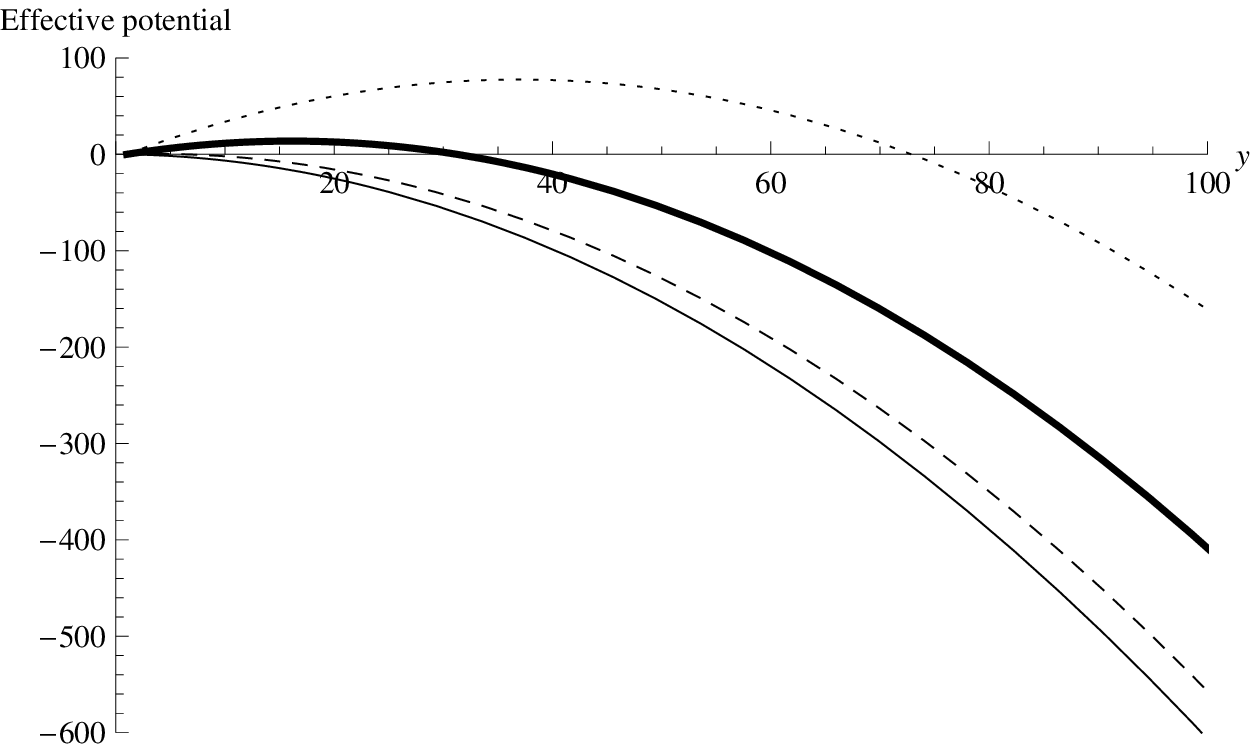}
\caption{The effective potential as a function of $y$ for $\xi=0$ (solid curve), $\xi=1$ (dashed curve), $\xi=2$ (thick curve),  $\xi=3$ (dotted curve), $l=1$, $m=1.8$ and $m=1.8i$, respectively.} 
\label{potential}
\end{figure}

To illustrate this, we consider a real scalar field in the Euclidean version
of metric (1) by performing a Wick rotation $\tau =it$, as in Ref.~\cite{Kachru:2008yh}, and considering ingoing waves at the horizon, the bulk
scalar field takes the asymptotic form 
\begin{equation}
\psi \sim c_{1}x^{\Delta _{+}}\psi _{+}\left( \tau ,\theta ,\phi \right)
+c_{2}x^{\Delta _{-}}\psi _{-}\left( \tau ,\theta ,\phi \right) ~,
\end{equation}
where $x=1-z$ and $\psi _{\pm }\left( \tau ,\theta ,\phi \right) $ are
sources for a dual field. $\Delta _{\pm }$ are roots of $\Delta \left(
\Delta -4\right) =m^{2}l^{2}$~, given by 
\begin{equation}
\Delta _{\pm }=2\left( 1\pm \frac{1}{2}\sqrt{4+m^{2}l^{2}}\right) =2\beta
_{\pm }~.
\end{equation}%
The on-shell Euclidean scalar field action is finite for $m^{2}l^{2}\geq -4$ 
\cite{Kachru:2008yh} and  \cite{Giacomini:2012hg}. Such as has been shown in Ref.~\cite{Kachru:2008yh}, for $-4<m^{2}l^{2}<-3$ there are two set of dual operators
with dimensions $\Delta _{+}$ and $\Delta _{-}$, respectively. Therefore, the
second set of quasinormal modes $\omega _{2}$ matches the dual operators
with $\Delta =\Delta _{-}$ for this range of scalar field mass. Now, we consider the tortoise coordinate 
\begin{equation}
\frac{dy^\ast}{dy}=\frac{l}{y\left( y-1\right) }~, 
\end{equation}
to analyze the behaviour of the effective potential at infinity. In this way, the radial equation~(\ref{equationy}),  can be written as a Schr\"{o}dinger equation
\begin{equation}\label{potential11}
\left( \frac{d^{2}}{dy^{\ast 2}}+\omega ^{2}-V\left( y\right) \right) \left(
y^{1/2}R\left( y^\ast \right) \right) =0~, 
\end{equation}
where
\begin{equation}\label{potential1}
V\left( y\right) =\left( 1-y\right) \left( \frac{1-y}{4l^{2}}-\frac{y}{2l^{2}%
}-\frac{Q}{2l^{2}}-\frac{m^{2}y}{4}\right)~. 
\end{equation}
We observe that the effective potential go to $+\infty$ for $3+m^2l^2>0$ and go to $-\infty$ for $3+m^2l^2<0$. See Fig.~(\ref{potential}).  
It is interesting note
that in the asymptotic limit $y\rightarrow \infty $ Eq. (\ref{potential11}) can be written
as a Schr\"{o}dinger equation with an inverse square potential given by 
\begin{equation}
V\left( y^{\ast }\right) \sim \left( \frac{m^{2}l^{2}}{4}+\frac{3}{4}\right) 
\frac{1}{y^{\ast 2}}~,
\end{equation}%
where the stability of the scalar field is given by $\frac{m^{2}l^{2}}{4}+%
\frac{3}{4}\geq -\frac{1}{4}$ \cite{Case:1950an} or $m^{2}l^{2}\geq -4$. The
same analysis has been performed for Lifshitz black hole in New Massive
Gravity,  \cite{Moon:2012dy}, where the behavior is similar.
Thus, we consider that the flux given by 
\begin{equation}\label{flux}
F=\frac{\sqrt{-g}g^{rr}}{2i}\left(R^*\partial_rR-R\partial_rR^*\right)~,
\end{equation} 
must vanishes at infinity. So, using~(\ref{R}), we obtain that the flux is 
\begin{eqnarray}
 F=\zeta \left\vert C_{1}\right\vert ^{2} Im\left( 2\left( \beta -1\right) B_{1}^{\ast }B_{2} +\left( \alpha +\gamma \right) \left\vert B_{1}\right\vert ^{2}\left( 1-z\right) ^{2\beta -1} +\left( \alpha +\eta \right) \left\vert B_{2}\right\vert ^{2}\left( 1-z\right) ^{3-2\beta}\right)~,
\end{eqnarray}
where
\begin{equation}
\zeta =\frac{l\sinh \theta }{2}~,
\end{equation}
\begin{equation}
\gamma=\frac{ab}{a+b-c}~,
\end{equation}
\begin{equation} 
\eta=\frac{(c-a)(c-b)}{c-a-b+1}~,
\end{equation}
\begin{equation}
B_1=\frac{\Gamma\left(c\right)\Gamma\left(c-a-b\right)}{\Gamma\left(c-a\right)\Gamma\left(c-b\right)}~,
\end{equation}
\begin{equation}
B_2=\frac{\Gamma\left(c\right)\Gamma\left(a+b-c\right)}{\Gamma\left(a\right)\Gamma\left(b\right)}~.
\end{equation}

If $ -4 \leq  m^2l^2 \leq 0$, then $1<\beta _{+}<2$ and $0<\beta _{-}<1$. So, $2\beta -1<0$ for $\beta =\beta _{-}$ and $3-2\beta <0$ for $\beta =\beta _{+}$. 
For,  $\beta_-$ the flux at infinity vanishes 
if $(c-a)|_{\beta_-}=-n$ or $(c-b)|_{\beta_-}=-n$. For $\beta_+$ the flux at infinity vanishes 
if $(a)|_{\beta_+}=-n$ or $(b)|_{\beta_+}=-n$. These conditions
determine the quasinormal modes. In spite of this, these conditions given the same QNMs that we have found (\ref{w1}) previously.  

For, $\beta=\beta_+$,  the conditions $2\beta_+ -1>0$ and $3-2\beta_+ >0$ are satisfied if $\sqrt{3}<Im(ml)<2$. In this case, the flux at infinity vanishes if $B_1=0$, i.e.,
if $(c-a)|_{\beta_+}=-n$ or $(c-b)|_{\beta_+}=-n$. Therefore, the quasinormal modes are
\begin{eqnarray}\label{w2}
\omega_2=-i\frac{2n-\sqrt{4+m^2l^2}}{4l}-i\frac{2Q-\sqrt{4+m^2l^2}+2n}{4l\left(-\sqrt{4+m^2l^2}+2n+1\right)}~.
\end{eqnarray}

The restriction $B_2=0$ yields the same QNMs we have previously found~(\ref{w1}). In this case, the imaginary part of the QNMs~(\ref{w2}) is negative if  $\sqrt{3}<Im(ml)<2$ for $Q\geq1/2$ and the imaginary part is negative if $\sqrt{2+2Q+2\sqrt{1-2Q}}<Im(ml)<2$ for $Q<1/2$. Therefore, under these conditions the four dimensional Topological Lifshitz black hole is stable. 

\section{Reflection coefficient, transmission coefficient and absorption cross section}
\label{coefficients}
The reflection and transmission coefficients depend on the behaviour
of the radial function both, at the horizon and at the asymptotic
infinity and they are defined by
\begin{equation}\label{reflectiond}\
R :=\left|\frac{F_{\mbox{\tiny asymp}}^{\mbox{\tiny
out}}}{F_{\mbox{\tiny asymp}}^{\mbox{\tiny in}}}\right|; \; T:=\left|\frac{F_{\mbox{\tiny
hor}}^{\mbox{\tiny in}}}{F_{\mbox{\tiny asymp}}^{\mbox{\tiny
in}}}\right|~,
\end{equation}
where $F$ is the flux (\ref{flux}). 
The behaviour at the horizon is given by~(\ref{Rhorizon}), and using~(\ref{flux}), we get the
flux at the horizon up to an irrelevant factor from the angular
part of the solution
\begin{equation}
\textit{F}
_{hor}^{in}=-\frac{1}{2}\omega l^{2} |C_{1}|^{2}~.
\end{equation}
To obtain the asymptotic behaviour of $R(z)$, we use $x=1-z$ in~(\ref{radial1}) and we consider the asymptotic infinity, i.e. $x\rightarrow 0$. Thus, we obtain the following equation
\begin{eqnarray}\label{equationx}
 x^2R''(x)-xR'(x)+\left(-\frac{m^2l^2}{4}+x\left(\frac{m^2l^2}{4}-\frac{Q}{2}\right)\right)R(x)=0~.
\end{eqnarray}
Whose solution is
\begin{eqnarray}
R_{asymp.}(x)= E_1\;B\;\Gamma\left(1-A\right)J_{-A}\left(2\sqrt{Bx}\right)x+E_2\;B\;\Gamma\left(1+A\right)J_{A}\left(2\sqrt{Bx}\right)x~,
\end{eqnarray}
where
\begin{equation}
A=2\sqrt{1+\frac{m^2l^2}{4}}~,\ \ B=\frac{m^2l^2}{4}-\frac{Q}{2}~,
\end{equation}
and using the expansion of the Bessel function \cite{M. Abramowitz}
\begin{equation}
J_n(q)=\frac{q^n}{2^n\Gamma{(n+1)}}\left\{1-\frac{q^2}{2(2n+2)}+...\right\},
\end{equation}
for $q\ll1$, we find the asymptotic infinity solution in the polynomial
form
 \begin{eqnarray}\label{Rasymp4}\
\nonumber R_{asymp.}(x)=E_1\;B^{1-A/2}\;x^{1-A/2} &&\\
+E_2\;B^{1+A/2}\;x^{1+A/2},
\end{eqnarray}
for $2\sqrt{Bx}\ll1$. Note that $1-A/2=\beta_-$ and~(\ref{Rasymp4}) match with~(\ref{R2}) with
\begin{equation}
E_1\;B^{1-A/2} = C_1 \frac{\Gamma(c)\Gamma(c-a-b)}{\Gamma(c-a)\Gamma(c-b)}=C_1B_1~, \end{equation}
\begin{equation}
E_2\;B^{1+A/2}= C_1 \frac{\Gamma(c)\Gamma(a+b-c)}{\Gamma(a)\Gamma(b)}=C_1B_2~.
\end{equation}   

The radial function~(\ref{Rasymp4}) is regular at infinity if $1-\frac{A}{2}\ge0$ or
$-4\leq m^2l^2\leq 0$, 
and $A$ cannot be an integer, due to $a+b-c=-A$, for
$\beta=\beta_-$, and $c-a-b=-A$, for $\beta=\beta_+$ and the gamma function is
singular at that point. The first condition is in agreement with the analogue to Breitenlohner-Freedman condition. This is, that any effective mass must be satisfied in order to have a stable propagation,~\cite{Breitenlohner:1982bm} and \cite{Breitenlohner:1982jf}.

Therefore, the flux~(\ref{flux}) at the asymptotic region~(\ref{Rasymp4}) is given by
\begin{equation}\label{fluxdinfinity}\
F_{asymp}\approx -\left(1-\beta \right)l\left|C_1\right|^2\emph{Im}[B_{2}B_{1}^{\ast }]~.
\end{equation}
up to an irrelevant factor from the angular part of the solution. 
In order to characterize the fluxes, we split up the coefficients
$B_1$ and $B_2$ in terms of the incoming and
outgoing coefficients, $B_{\mbox{\tiny in}}$ and $B_{\mbox{\tiny
out}}$, respectively. We define  $B_1 = B_{\mbox{\tiny
in}} + B_{\mbox{\tiny out}}$ and $B_2 = i h
(B_{\mbox{\tiny out}} - B_{\mbox{\tiny in}})$ with $h$ being a
dimensionless constant which it will be assumed to be independent of
the energy $\omega$ \cite{Birmingham:1997rj}, \cite{Kim:1999un}, \cite{Oh:2008tc},
\cite{Kao:2009fh}, \cite{Gonzalez:2010ht}, \cite{Gonzalez:2010vv} and \cite{Lepe:2012zf}.  In this way, we can write the asymptotic flux~(\ref{fluxdinfinity}) as
\begin{equation}
F_{asymp}\approx (1-\beta )lh\left|C_1\right|^2[|B_{in}|^{2}-|B_{out}|^{2}]~.
\end{equation}
Therefore, the reflection and transmission coefficients
 are given by
\begin{equation}
R=\frac{|B_{out}|^{2}}{|B_{in}|^{2}}
~,\label{coef1}
\end{equation}
\begin{equation}
T=\frac{\omega l}{2|1-\beta ||h||B_{in}|^{2}}
~,\label{coef2}
\end{equation}
and the absorption cross section $\sigma_{abs}$, becomes
\begin{equation}\label{absorptioncrosssection}\
\sigma_{abs}=\frac{ l}{2|1-\beta ||h||B_{in}|^{2}}~,
\end{equation}
where, the coefficients $|B_{\mbox{\tiny in}}|^2$ and $|B_{\mbox{\tiny
out}}|^2$ are given by
\begin{equation}\label{D11}\
 |B_{\mbox{\tiny in}}|^2= \frac{1}{4}\{|B_{1}|^{2}+\frac{1}{h^{2}}|B_{2}|^{2}-\frac{2}{h}%
\emph{Im}[B_{2}B_{1}^{\ast }]\}~,
 \end{equation}
\begin{equation}\label{D21}\
| B_{\mbox{\tiny out}}|^2=
\frac{1}{4}\{|B_{1}|^{2}+\frac{1}{h^{2}}|B_{2}|^{2}+\frac{2}{h}%
\emph{Im}[B_{2}B_{1}^{\ast }]\}~.
 \end{equation}
 
The poles of the
transmission coefficient are given by $\left\vert B_{in}\right\vert ^{2}=0$.
Using equation (\ref{D11}) and solving for $h$, we find the following relations
must hold for the existence of a pole
\begin{equation}
h=\frac{Im\left( B_{1}^{\ast }B_{2}\right) }{\left\vert
B_{1}\right\vert ^{2}},\ \ \ \ Im\left( B_{1}^{\ast
}B_{2}\right) =\pm \left\vert B_{1}\right\vert \left\vert B_{2}\right\vert ~.
\end{equation}
Solving these equations for $\omega $ it could be possible to obtain a set
of frequencies, however it is a difficult task. An interesting question is
to see if these poles coincide with the quasinormal modes $\omega _{1}$
obtained in section II. Due that $\omega _{1}$ are purely imaginary, the
constants $\alpha $, $\beta $, $a$, $b$, $c$, $B_{1}$ and $B_{2}$ are real
when evaluated in these modes, therefore $Im\left( B_{1}^{\ast
}B_{2}\right) =0$, and from equation (\ref{D11}) we see that for these QNMs to
coincide with the poles of the transmission coefficient, the following
relation must hold 
\begin{equation}
\left\vert B_{1}\right\vert ^{2}+\frac{1}{h^{2}}\left\vert B_{2}\right\vert
^{2}=0~,
\end{equation}
but this relation implies $\left\vert B_{1}\right\vert =\left\vert
B_{2}\right\vert =0$. However $\left\vert B_{1}\right\vert $ and $\left\vert
B_{2}\right\vert $ can't be both zero. Therefore the QNMs $\omega _{1}$
doesn't correspond to the poles of the transmission coefficient. 
 
\begin{figure}[h]
\begin{center}
\includegraphics[width=0.45\textwidth]{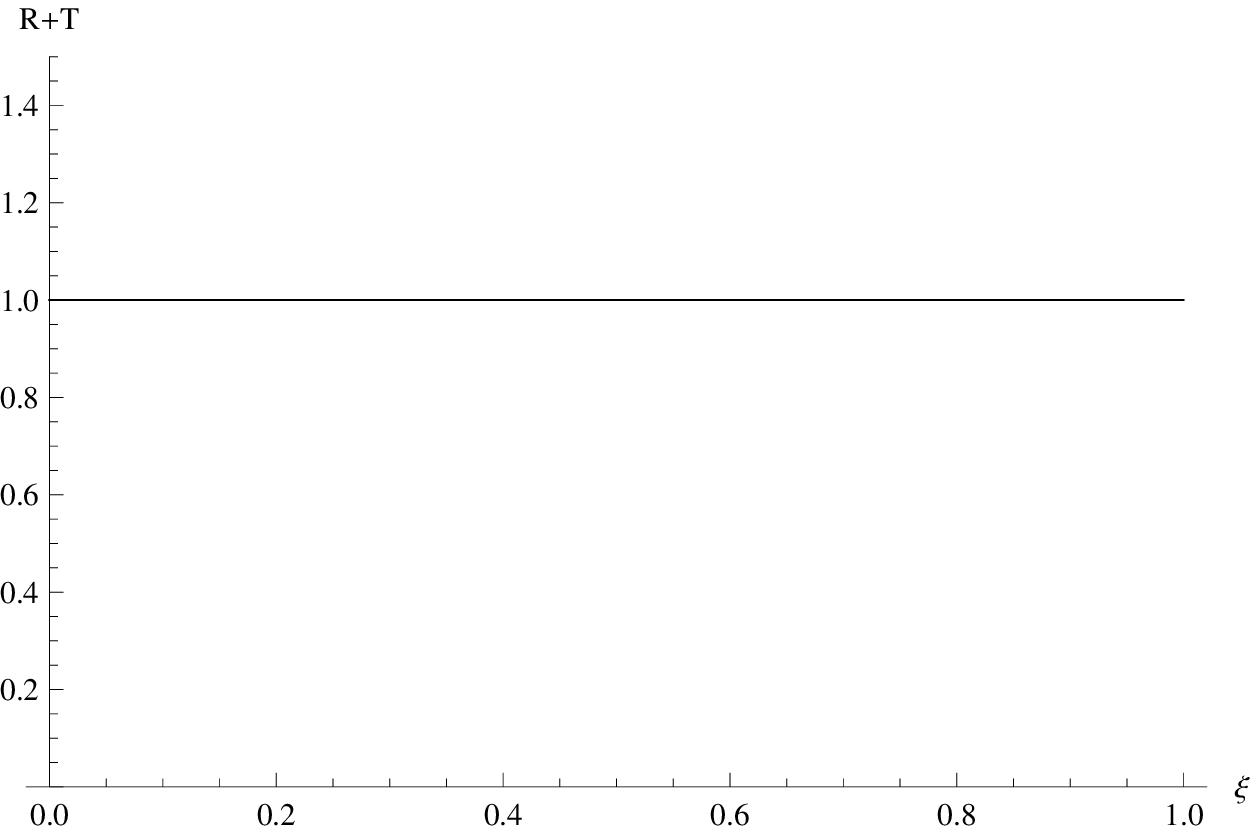}
\includegraphics[width=0.45\textwidth]{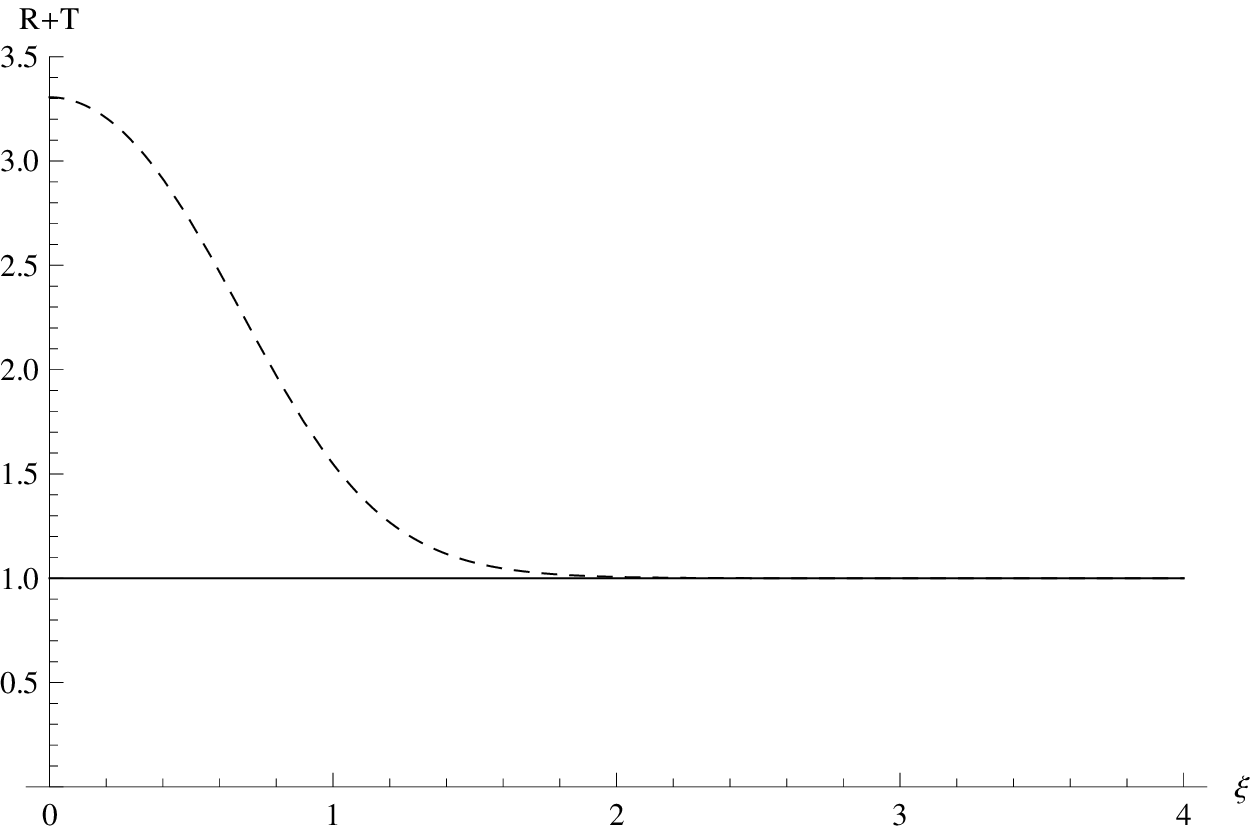}
\end{center}
\caption{\textit{R+T as a function of $\xi $; for $h=-1$ and $h=1$, respectively. Solid curve for $\omega =0$
and dashed curve for $\omega =1$; $m=i$ and $l=1$.}}
\label{RT}
\end{figure}

\begin{figure}[h]
\begin{center}
\includegraphics[width=0.45\textwidth]{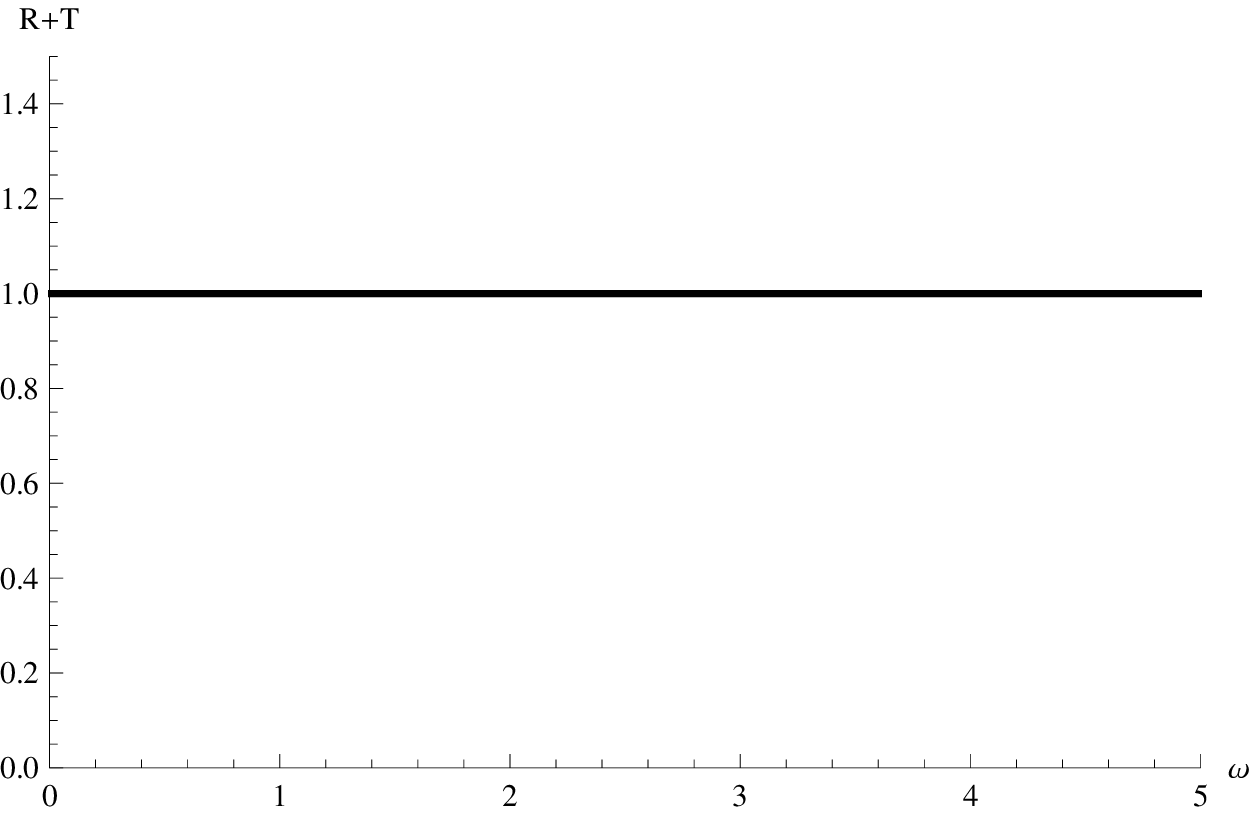}
\includegraphics[width=0.45\textwidth]{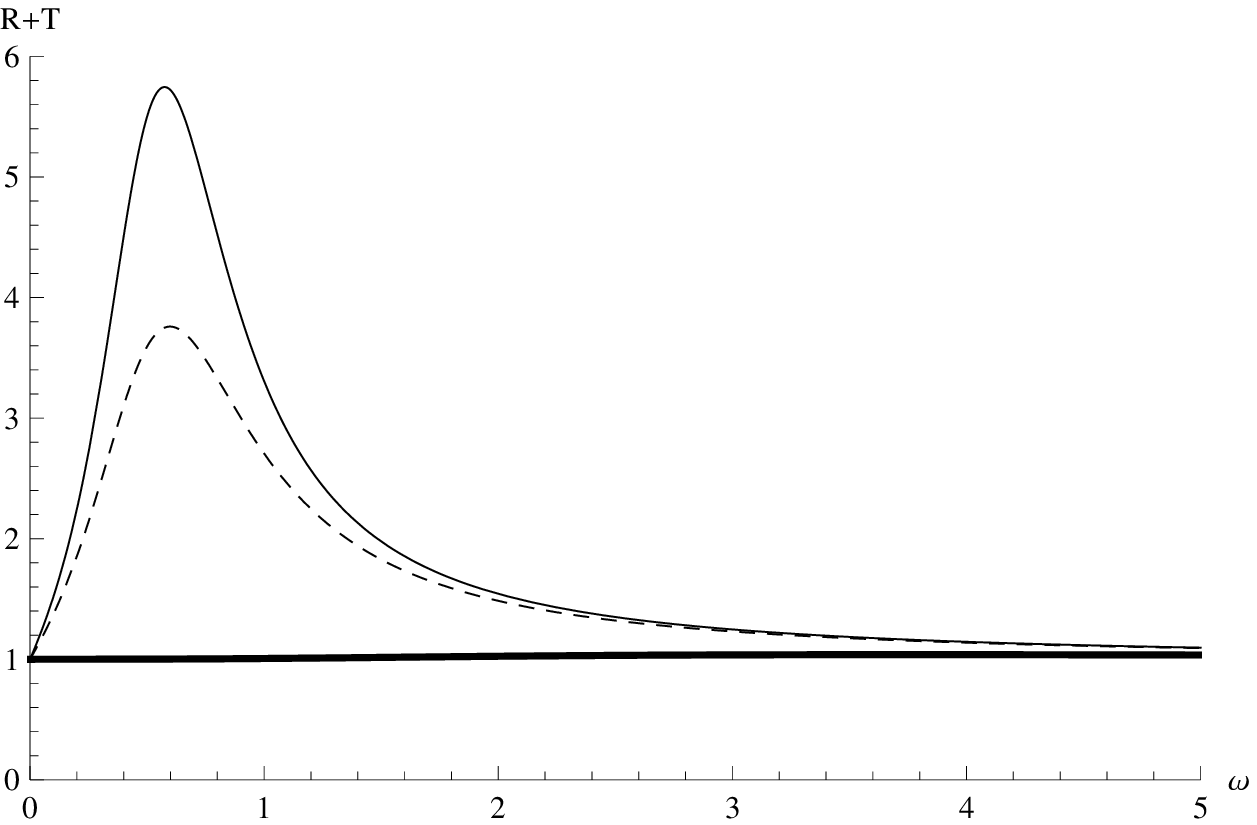}
\caption{{\it{$R+T$ as a function of $\omega$, for $h=-1$ and $h=1$, respectively. Solid curve for $\xi=0$, dashed curve for $\xi =0.5$ and thick curve for $\xi =2$; $m=i$ and $l=1$.}}}
\label{CaseRTomega}
\end{center}
\end{figure}

It is known that the constant $h$ can be chosen so that the absorption cross section can be
expressed by the area of horizon in the zero-frequency limit \cite{Birmingham:1997rj} and \cite{Das:1996we}. Also, it can be chosen to obtain the usual value of the Hawking temperature \cite{Kim:1999un} or in such a way so that the sum of the reflection coefficient and the transmission coefficient be unity \cite {Kim:2004sf}. For negative $h$ we obtain that $R+T=1$ and for positive $h$
this relation doesn't always hold, see Fig.~(\ref{RT}) and (\ref{CaseRTomega}). 
Based on this, in this
work we take arbitrarily, but without loss of generality $h=-1$ in all the
pictures in such way to ensure that $R+T=1$ for any angular momentum, the
qualitative behavior of the coefficients is the same for any negative $h$. 

\begin{figure}
\begin{center}
\includegraphics[width=0.45\textwidth]{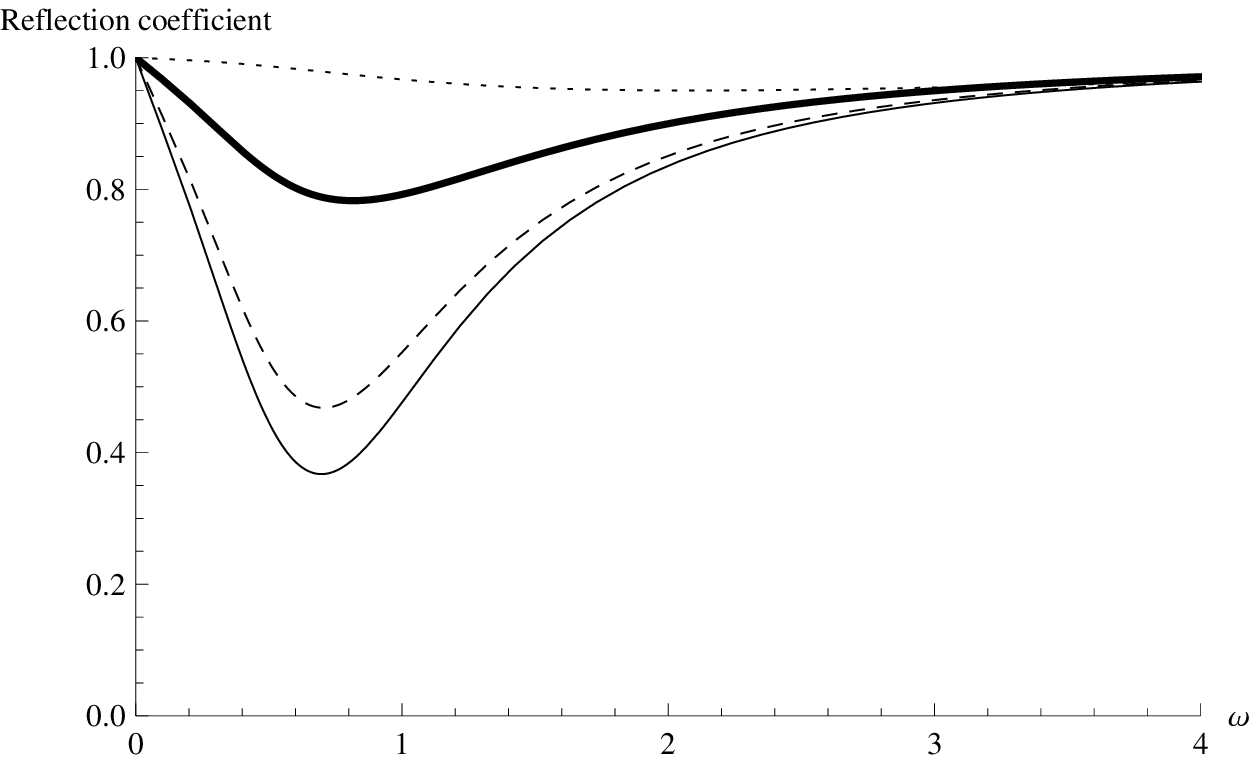}
\includegraphics[width=0.45\textwidth]{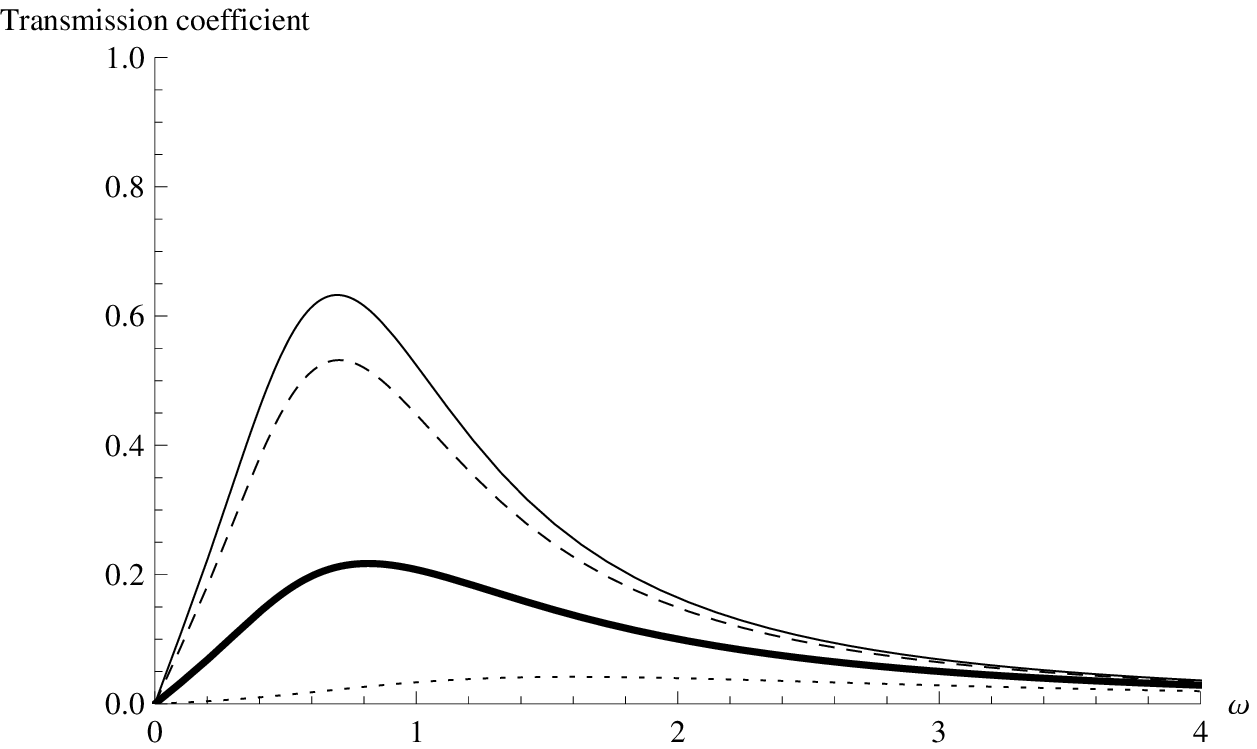}
\includegraphics[width=0.45\textwidth]{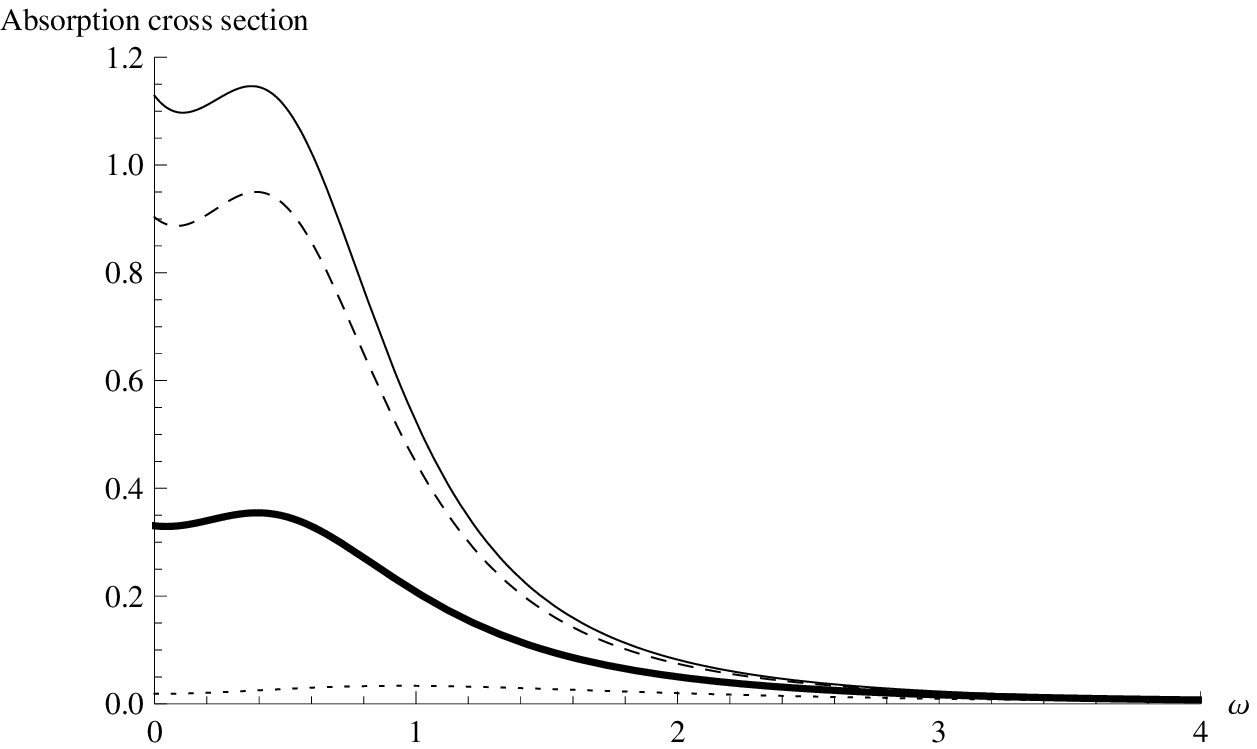}
\caption{The reflection coefficient, the transmission coefficient and the absorption cross section as a function of $\omega$; for $\xi=0$ (solid curve), $\xi=0.5$ (dashed curve), $\xi=1$ (thick curve),  $\xi=1.5$ (dotted curve),
$m=1$, $l=1$ and $h=-1$.}
\label{Analysis1}
\end{center}
\end{figure}

\begin{figure}
\begin{center}
\includegraphics[width=0.45\textwidth]{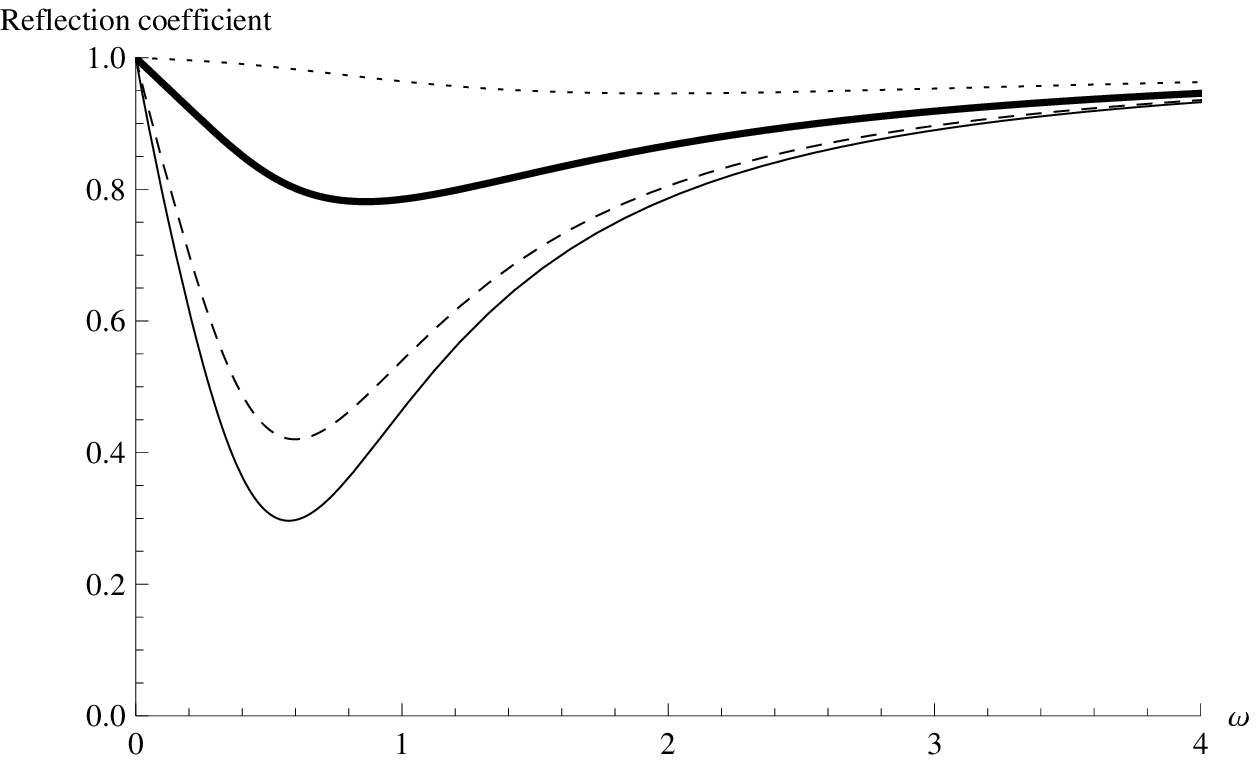}
\includegraphics[width=0.45\textwidth]{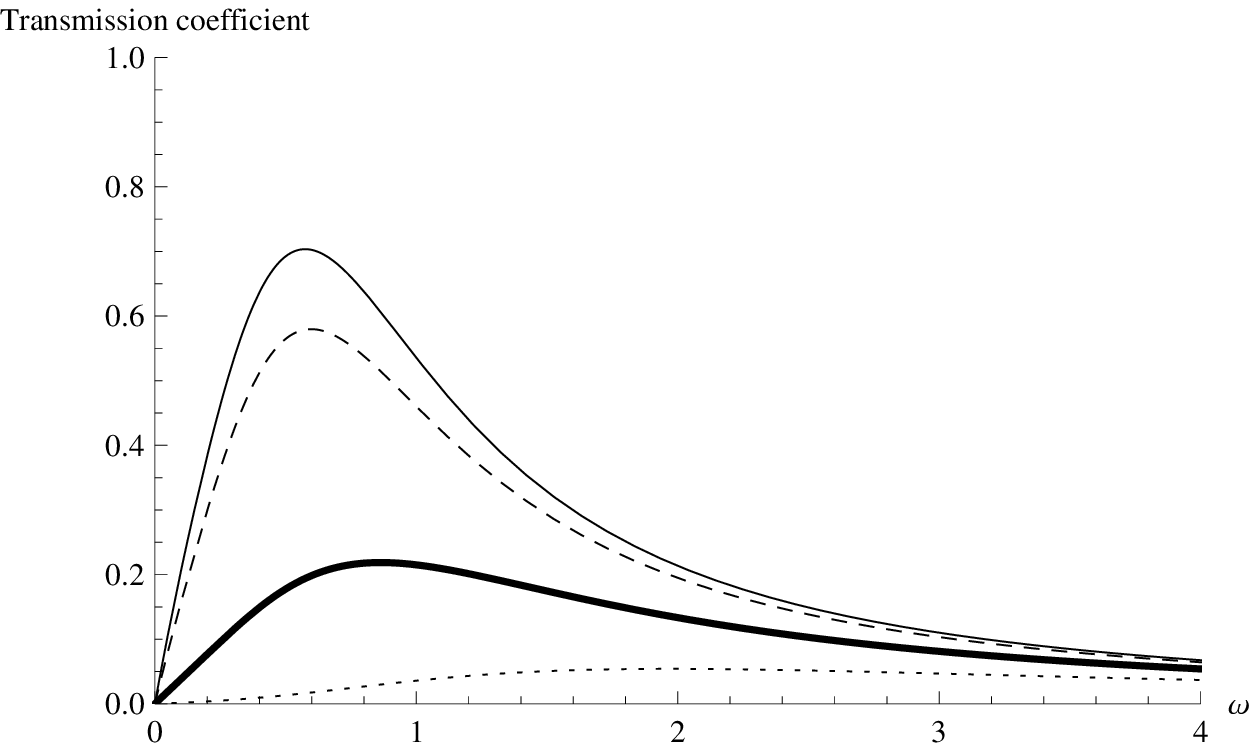}
\includegraphics[width=0.45\textwidth]{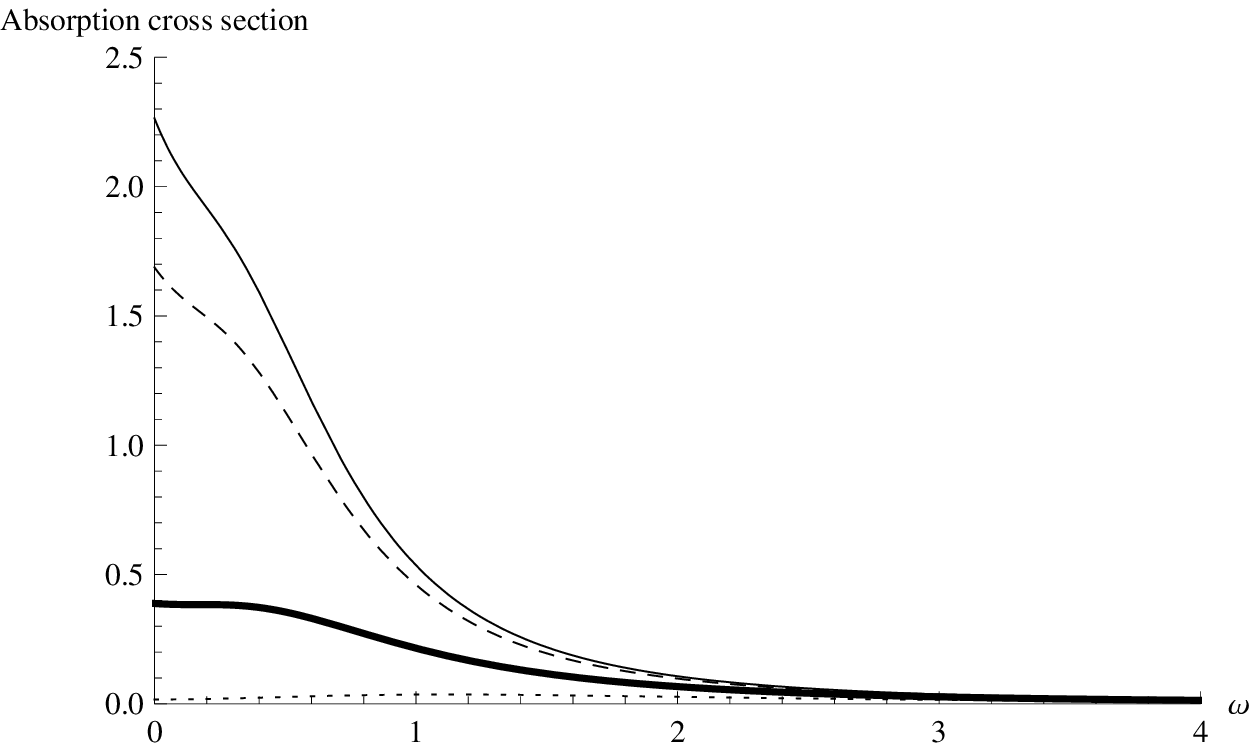}
\caption{The reflection coefficient, the transmission coefficient and the absorption cross section as a function of $\omega$; for $\xi=0$ (solid curve), $\xi=0.5$ (dashed curve), $\xi=1$ (thick curve),  $\xi=1.5$ (dotted curve),
$m=i$, $l=1$ and $h=-1$.}
\label{Analysis2}
\end{center}
\end{figure}

\begin{figure}[h]
\begin{center}
\includegraphics[width=0.45\textwidth]{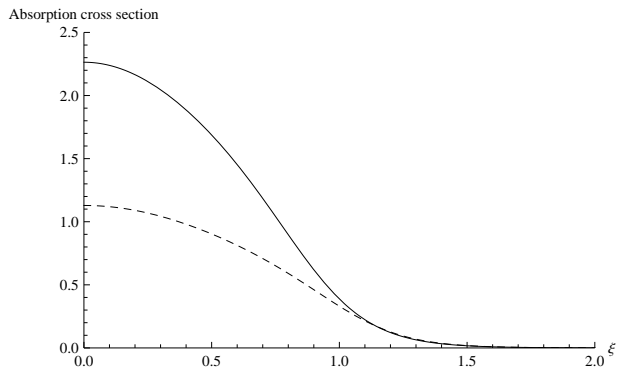} 
\end{center}
\caption{\textit{Absorption cross section as a function of $\protect\xi $;
for $m=i$ (solid curve), $m=1$ (dashed curve), $l=1$, $\protect\omega =0$
and $h=-1$.}}
\label{Absorption}
\end{figure}

\begin{figure}[h]
\begin{center}
\includegraphics[width=0.45\textwidth]{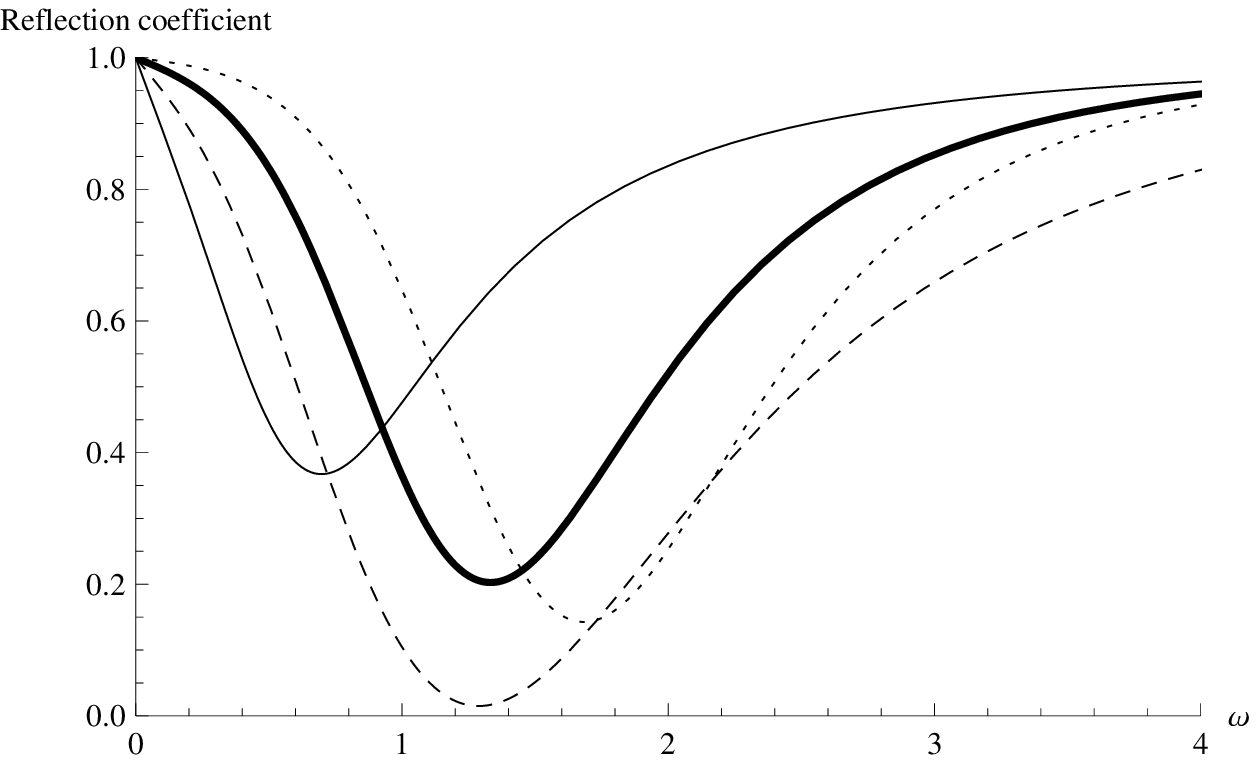}
\includegraphics[width=0.45\textwidth]{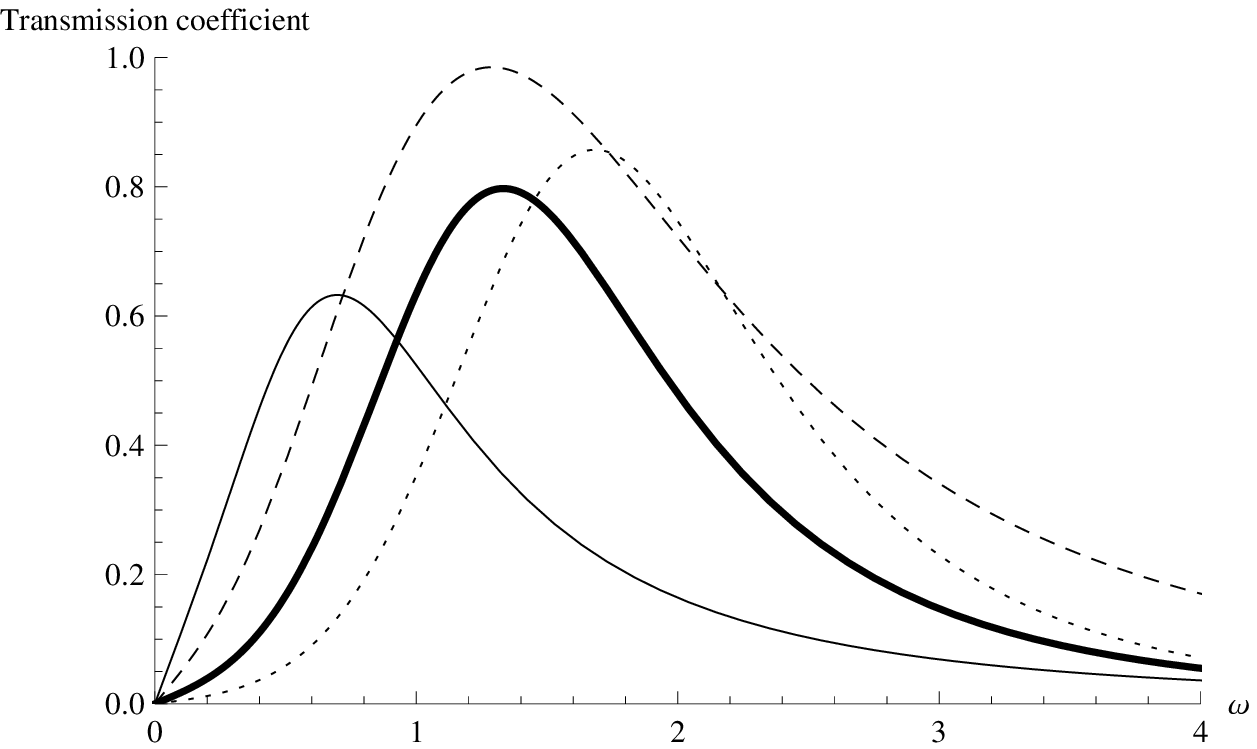}
\includegraphics[width=0.45\textwidth]{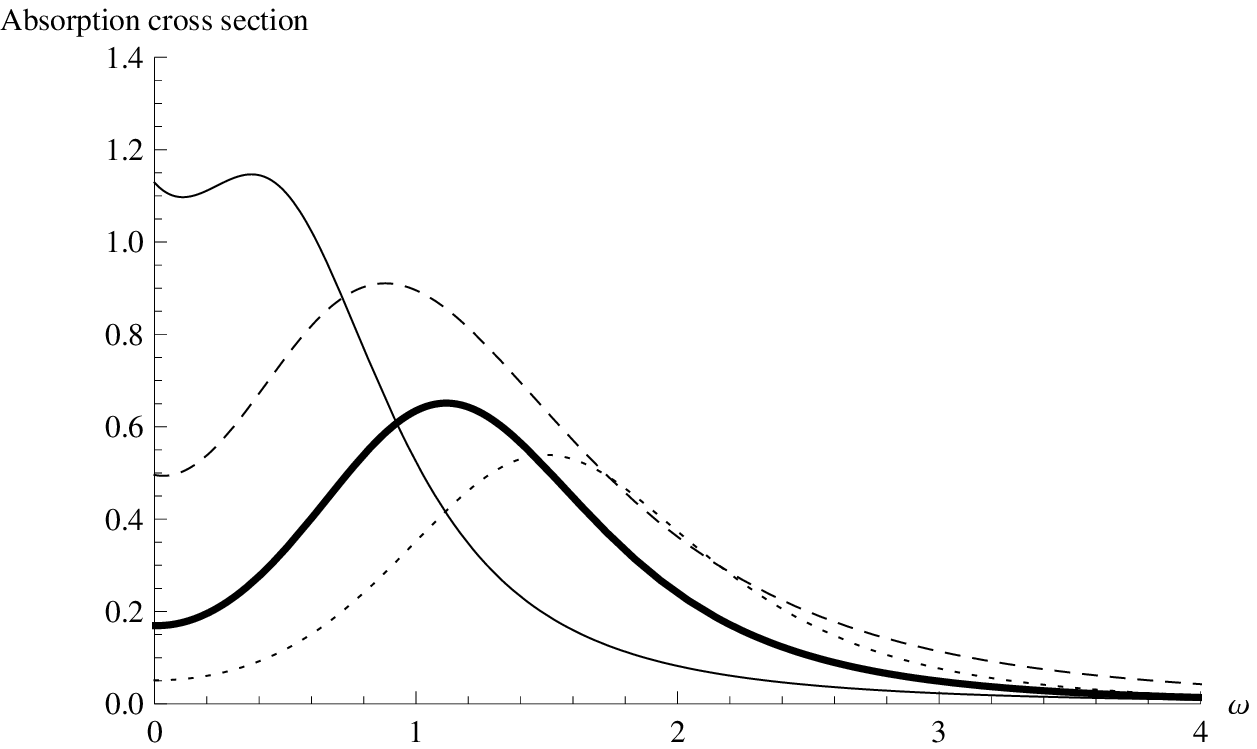}
\caption{The reflection coefficient, the transmission coefficient and the absorption cross section as a function of $\omega$; for $m=1$ (solid curve), $m=2$ (dashed curve), $m=3$ (thick curve),  $m=4$ (dotted curve), $\xi=0$, $l=1$ and $h=-1$.}
\label{Analysis1m}
\end{center}
\end{figure}

\begin{figure}[h]
\begin{center}
\includegraphics[width=0.45\textwidth]{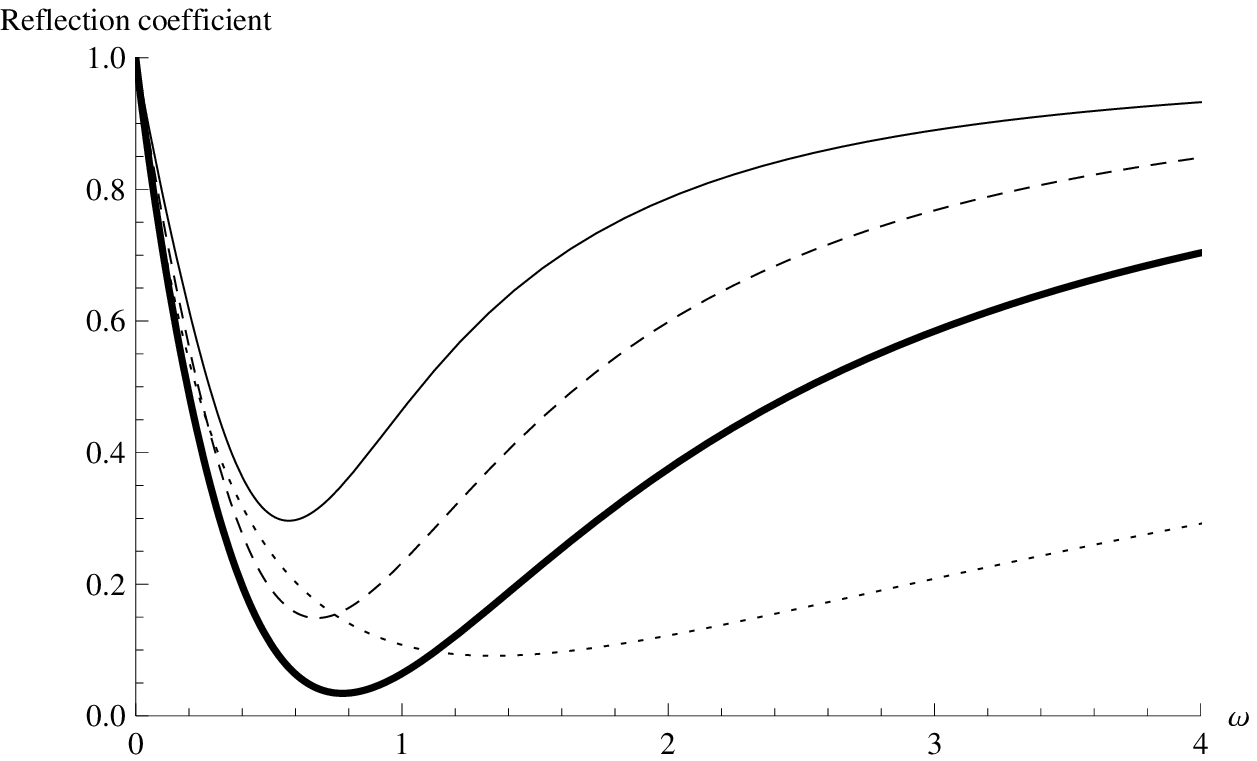}
\includegraphics[width=0.45\textwidth]{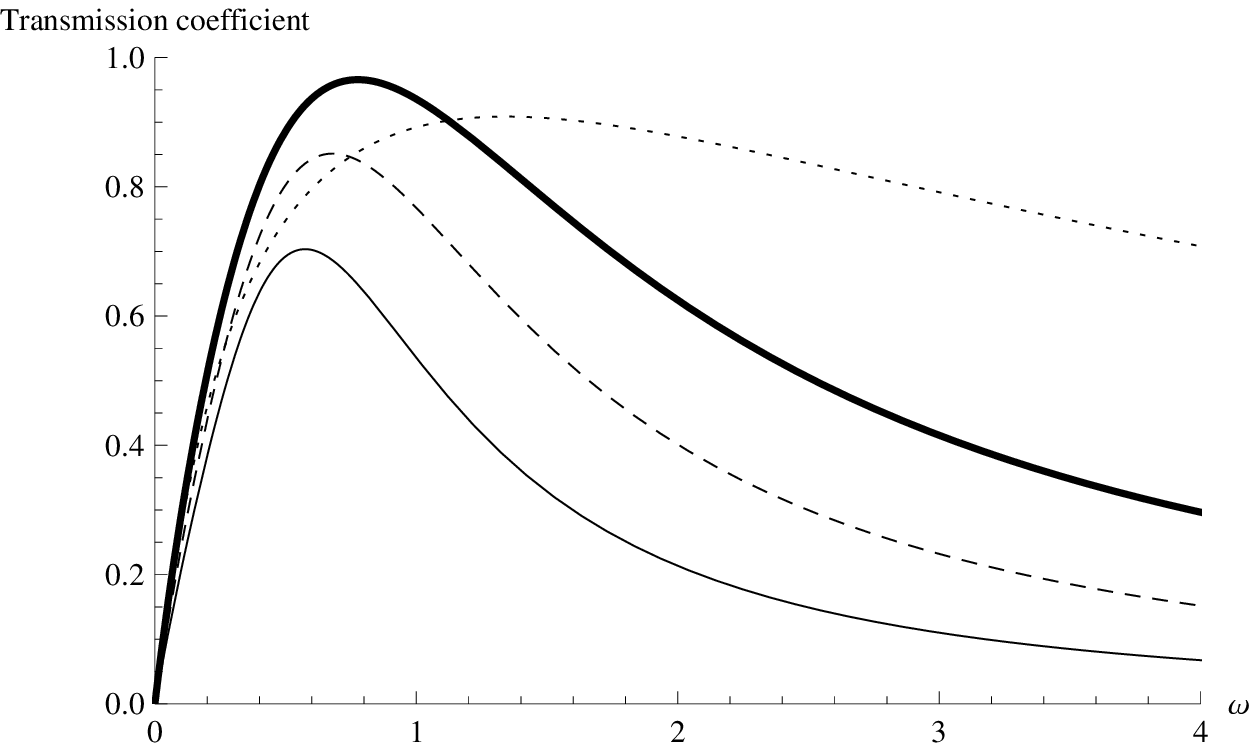}
\includegraphics[width=0.45\textwidth]{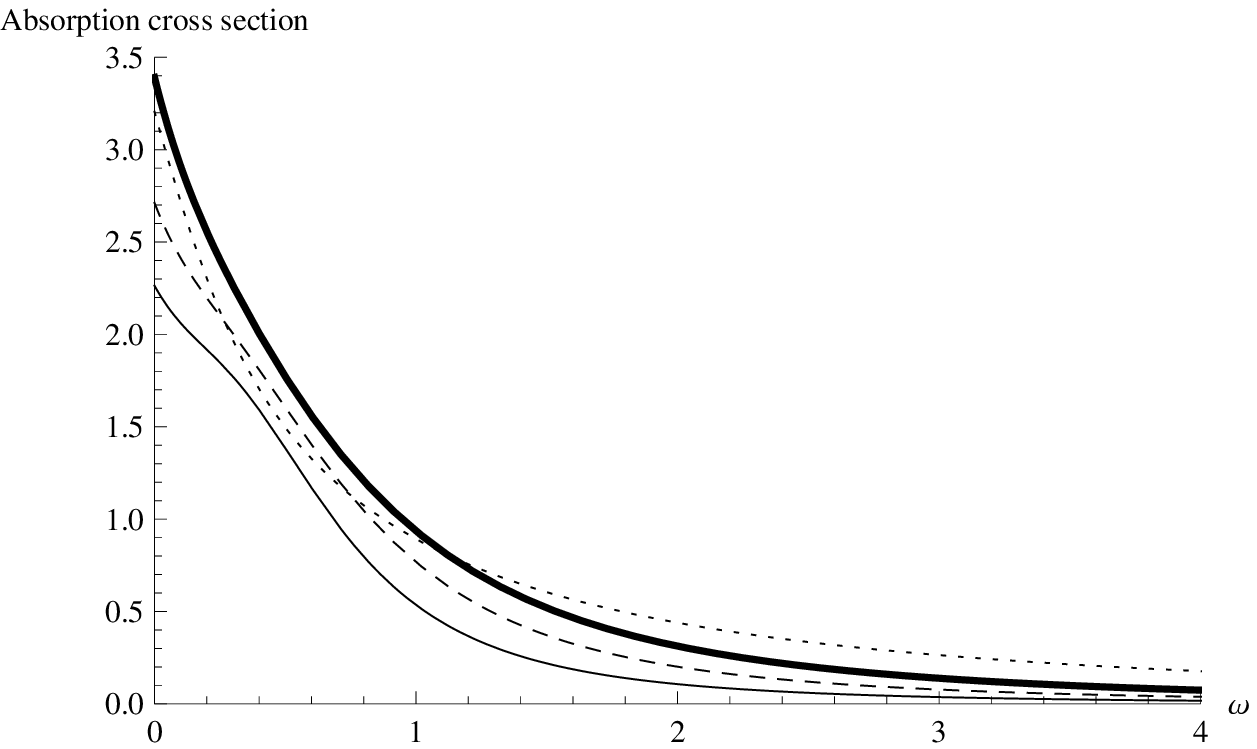}
\caption{The reflection coefficient, the transmission coefficient and the absorption cross section as a function of $\omega$; for $m=i$ (solid curve), $m=1.2i$ (dashed curve), $m=1.4i$ (thick curve),  $m=1.8i$ (dotted curve), $\xi=0$, $l=1$ and $h=-1$.}
\label{Analysis2m}
\end{center}
\end{figure}

Now, we will carry out a  numerical analysis of the reflection~(\ref{coef1}), and transmission coefficients~(\ref{coef2}), for a four dimensional topological Lifshitz black hole.
We consider, without loss of generality, $l=1$ and we fix $h=-1$. Note that in~(\ref{D11}) if $h$ is negative, the coefficients are regular and then we plot them for $\xi=0,0.5,1,1.5$. See Figs.~(\ref{Analysis1}, \ref{Analysis2}), for the reflection, transmission coefficients and the greybody factors, for positive scalar field mass ($m=1$) and imaginary scalar field mass ($m=i$), respectively. Here, we observe that in the zero-frequency limit, there is a range of values of $\xi$ which contribute to the absorption cross section, see Fig.~(\ref{Absorption}), as it happens in \cite{Gonzalez:2010ht}, \cite{Gonzalez:2010vv}, \cite{Gonzalez:2011du} and \cite{Lepe:2012zf}. Also, we observed in the low frequency limit, that the reflection and
transmission coefficients show a minimum and a maximum. Therefore,
the coefficients have two branches in the reflection case,
decreasing for low frequencies and then increasing. In the
transmission case the behaviour is opposite, they are increasing
and then decreasing, in such a way that $R+T=1$ occurs in all
cases. 
In order to study the effect of the mass of the scalar field, we plot  the reflection and transmission
coefficient and the absorption cross section for the mode with lowest angular momentum $\xi=0$, with $l=1$ and $h=-1$ in Figs.~(\ref{Analysis1m}, \ref{Analysis2m}), and we consider positive scalar field mass ($m=1,2,3,4$) and imaginary scalar field mass ($m=i, 1.2i, 1.4i,1.8i$), respectively. It is worth to mention that the coefficients are regular for positive scalar field mass, but~(\ref{Rasymp4}) is not regular for positive scalar field mass. Essentially, we found the same general behaviour for the different values of $m$ and the only difference is a shift in the location of the minimum or maximum of the reflection and transmission coefficients, respectively. We observe that the absorption cross is decreasing if the parameter $m$ is increasing in the low frequency limit for real scalar field mass, Fig~(\ref{Analysis1m}). We also can observe the existence of one optimal frequency to transfer energy out of the bulk and the absorption cross section vanishes in the low frequency limit, for $\xi=0$ and $m\rightarrow 0$. 


\section{Conclusions}
\label{remarks}
The Lifshitz black hole is a very interesting static solution of gravity theories which asymptotically approach to Lifshitz spacetimes. 
In this work we have studied four dimensional Topological Lifshitz black hole under scalar perturbations. We have considered Dirichlet and Neumann boundary conditions and we have found that the quasinormal modes are purely imaginary i.e., there is only damped modes, which guaranteed the stability of Lifshitz black hole due to the imaginary part of the QNMs is negative. Also, we have computed analytically the reflection and transmission coefficients and the absorption cross section and we have found that there is a range of modes with high angular momentum which contributes to the absorption cross section in the low frequency limit. Furthermore,  in this limit, the absorption cross section decreases if the scalar field mass increase, for a real scalar field mass.

\section*{Acknowledgments}
Y. V. is supported by FONDECYT grant 11121148, and by Direcci\'{o}n de Investigaci\'{o}n y Desarrollo,
Universidad de la Frontera, DIUFRO DI11-0071.  P.G. acknowledges the hospitality of the Physics Department of Universidad de La Frontera where part of this work was made.

%
%
%
%
%

\end{document}